\newif\ifjfm\jfmfalse
\newif\ifaps\apstrue
\newif\ifsectioning\sectioningtrue
\newif\ifparanumbering\paranumberingfalse

\documentclass[english, aps, amsmath, amssymb, preprint]{revtex4-1} 

\usepackage[english]{babel}
\usepackage{graphicx}
\usepackage{verbatim}
\usepackage{hyperref}
\hypersetup{linkcolor=blue, citecolor=blue, urlcolor  = blue}
\ifaps
	\usepackage{color}
	\usepackage{breakurl}
\fi
\ifparanumbering
\fi

\graphicspath{ {./figures/} } 

\newcommand\manuscripttitle{The effect of mean flow speed in soap film channels}
\newcommand\authorAname{Ildoo \surname{Kim}}
\newcommand\authorAaddressA{Department of Mechatronics, Konkuk University, Chungju, South Korea 27478}
\newcommand\authorAaddressB{KU Advanced Artificial Intelligence Robot Research Center, Konkuk University, Chungju, South Korea 27478}
\newcommand\authorAemail{ildookim@kku.ac.kr}

\newcommand{\vm}{v_m}

\newcommand{\Reynolds}{{\textit{Re}}}

\ifaps
	\newcommand\citepar\cite
	\newcommand\citebyname\cite

	\makeatletter
\fi

\begin{document}

\ifaps
	\title{\manuscripttitle}
	\author{\authorAname}
	\email{\authorAemail}
	\affiliation{\authorAaddressA}
	\affiliation{\authorAaddressB}
	\date{\today}
\fi

\ifjfm
	\maketitle
\fi

\begin{abstract}
The soap film channel has been developed as a model system for two-dimensional hydrodynamics, but its general applicability has been questioned because the mean flow speed is known to alter the experimental outcome considerably, for reasons that have not been examined.
%It is widely acknowledged in the community that consistent measurements require the mean flow speed to be held fixed, yet the underlying mechanism has not been examined.
In this study, we experimentally investigate how the geometry of a vortex street in a flowing soap film varies with the mean flow speed, using both an inclined and a vertical channel.
We quantify the geometry by the Kármán ratio $q$, the ratio of the transverse to the longitudinal spacing between vortices, and find that the variation of $q$ is accounted for by none of the object size, flow speed $u$, or film thickness $\delta$ alone.
However, the data from both channels collapse onto a single curve when plotted against $u\sqrt{\delta}$, which is proportional to a Mach-like number $M\equiv u/\vm$, where $\vm$ is the Marangoni elastic wave speed.
The measured collapse follows $q/q_0=1-M^2$, an $\mathcal{O}(M^2)$ correction characteristic of weakly compressible flow.
The breakdown of Reynolds similitude in soap film channels therefore originates from the two-dimensional compressibility of the film.
\end{abstract}

\ifjfm
\fi

\ifaps
	\maketitle
\fi

%%%%%%%%%%%%%%%
\ifsectioning\section{Introduction}\fi
%%%%%%%%%%%%%%%

\ifparanumbering\paragraph{}\fi
Experiments in physics take place in a model system.
A model system is a simple and approximate version of reality, and it should be designed to capture key physics under investigation. 
Ideally, the system is well understood and controlled to avoid any unnecessary complications or uncertainties to have a reproducible and consistent result.

\ifparanumbering\paragraph{}\fi
This work concerns a soap film channel as a model system of two-dimensional (2D) fluid.
The soap film channel has been developed as a scientific instrument in late 1980s with seminal contributions 
%by \citebyname{Couder-pd-89}, \citebyname{Beizaie:1997to}, and \citebyname{Rutgers:2001wj}, 
by various groups \citepar{Couder-pd-89, Beizaie:1997to, Chomaz:2001us, Rutgers:2001wj}, 
and since then it has been used to study 2D hydrodynamics.
Specifically, soap film channels are used to investigate cylinder wakes \citepar{Vorobieff:1999wn,Roushan:2005un,Kim:2015jp,Kim:2019kw}, flow past elastic structures \citepar{Jung:2006tn,Ristroph:2008wh}, and 2D pipe flows \citepar{Tran:2010hn,Cerbus:2018jc}.
Most notably, great advances have been made in our understanding of 2D decaying and forced turbulence \citepar{Kellay:1995wk,Martin:1998ty}, including important observations of the inverse cascade \citepar{Rutgers:1998uz}.

\ifparanumbering\paragraph{}\fi
Despite numerous success stories, the soap film channel remains a capricious system that requires great caution to obtain consistent results.
It is widely acknowledged by the community that the variation of flow speed greatly alters the outcome of the experiments, and many research groups persistently keep the flow speed of their soap film channels invariant throughout a study.
It is speculated that variations in mean flow speed accompanies the variation in film thickness \citepar{Rutgers:1996vg,Sane:2018uo}, which causes the experimental uncertainty.
However, the exact physical mechanism remains unexplored.

\ifparanumbering\paragraph{}\fi
In this work, we investigate why and how the variations in the mean flow speed $u$, as well as the accompanied variation in film thickness $\delta$, affect the downstream flow structure.
The very fact that $u$ and $\delta$ are coupled suggests that the soap film is three-dimensionally incompressible but 2D-compressible, which brings us to consider the gasdynamic analog of soap film flows \citepar{Wen:2003td, Wen:2003wz, Fast:2005wq}.
With non-negligible Marangoni stress term $\nabla\sigma$, where $\sigma$ is the surface tension, the momentum equation of soap film flows becomes similar to that of compressible gas.
In particular, $\delta$ is analogous to the density of gas, and the Marangoni elastic wave speed $\vm$ is analogous to the acoustic wave speed (the speed of sound).
As the density fluctuation propagates at the acoustic wave speed, the fluctuation in $\delta$ propagates at $\vm$, which is expressed in the form
\begin{equation}
	\vm=\sqrt{\frac{2E}{\rho \delta}},
	\label{eq:Marangoni_wave_speed}
\end{equation}
where $\rho$ is the density of water and $E\equiv -\partial\sigma/\partial(\ln \delta)$ is the Marangoni elasticity \citepar{Lucassen:1970ta}.
The surfactants provide the mechanical stability to soap films, and a 2D-compressible wave is produced in this dynamic process \citepar{Taylor:1959ta,Lucassen:1970ta,Vrij:1970vs,Couder-pd-89}.
By observing oblique shocks in soap films, \citebyname{Kim:2017dn} measured the value of the Marangoni elasticity, $E\approx22 \,\rm mN/m$.
For a typical setup, $\vm$ ($\sim3.0\,\rm m/s$) is comparable to $u$ ($\sim 1.5 \,\rm m/s$).
% for a typical setup for soap film channels.
%Interestingly, $\vm$ is measured to be $\sim3.0\,\rm m/s$ depending on $\delta$, comparable to the characteristic flow speed ($\sim 1.5 \,\rm m/s$).

\ifparanumbering\paragraph{}\fi
This study has two features.
First, we use the Kármán ratio $q$ to quantify the geometric structure of vortex streets.
If the soap film flow is a 2D Navier-Stokes system, the principle of similitude should hold; when the flow is rescaled according to its characteristic length, velocity, and time scales, it should be indistinguishable from the another flow of the same Reynolds number.
However, we observe counter examples, which will be quantified by the Kármán ratio.
Second, we use two different soap film channels, an inclined and a vertical channel.
Generally speaking, the inclined channel produces a slower and thicker film than the vertical channel.
By using both, we vary $u$ and $\delta$ somewhat independently.

\ifparanumbering\paragraph{}\fi
Our experimental investigation find that $q$ is determined by none of the object size $D$, $u$, or $\delta$ alone but scales with $u\sqrt\delta$.
This parameter is the only possible combination of $u$ and $\delta$ via Buckingham Pi theorem, and physically, it can be interpreted as a Mach-like number $M\equiv u/\vm$.
Regression analysis shows that $q/q_0=1-u^2\delta/\Lambda^2$, where $\Lambda$ is the fitting parameter, indicating that the breakdown of Reynolds similitude in soap film flows originates from the compressibility of the film.

%%%%%%%%%%%%%%%
\ifsectioning\section{Experimental set-up}\label{sec2}\fi
%%%%%%%%%%%%%%%

\ifparanumbering\paragraph{}\fi
The experiments are carried out using an inclined and a vertical soap film channel.
The main difference between these two channels is the relationship between the film thickness $\delta$ and the mean flow speed $u$. 
In this study, we investigate the structural change in vortex streets by the variation of $\delta$ and $u$, therefore the use of two channels with different $\delta$-$u$ relationship is the key element.

\begin{figure}
\begin{centering}
\includegraphics[width=0.408\textwidth]{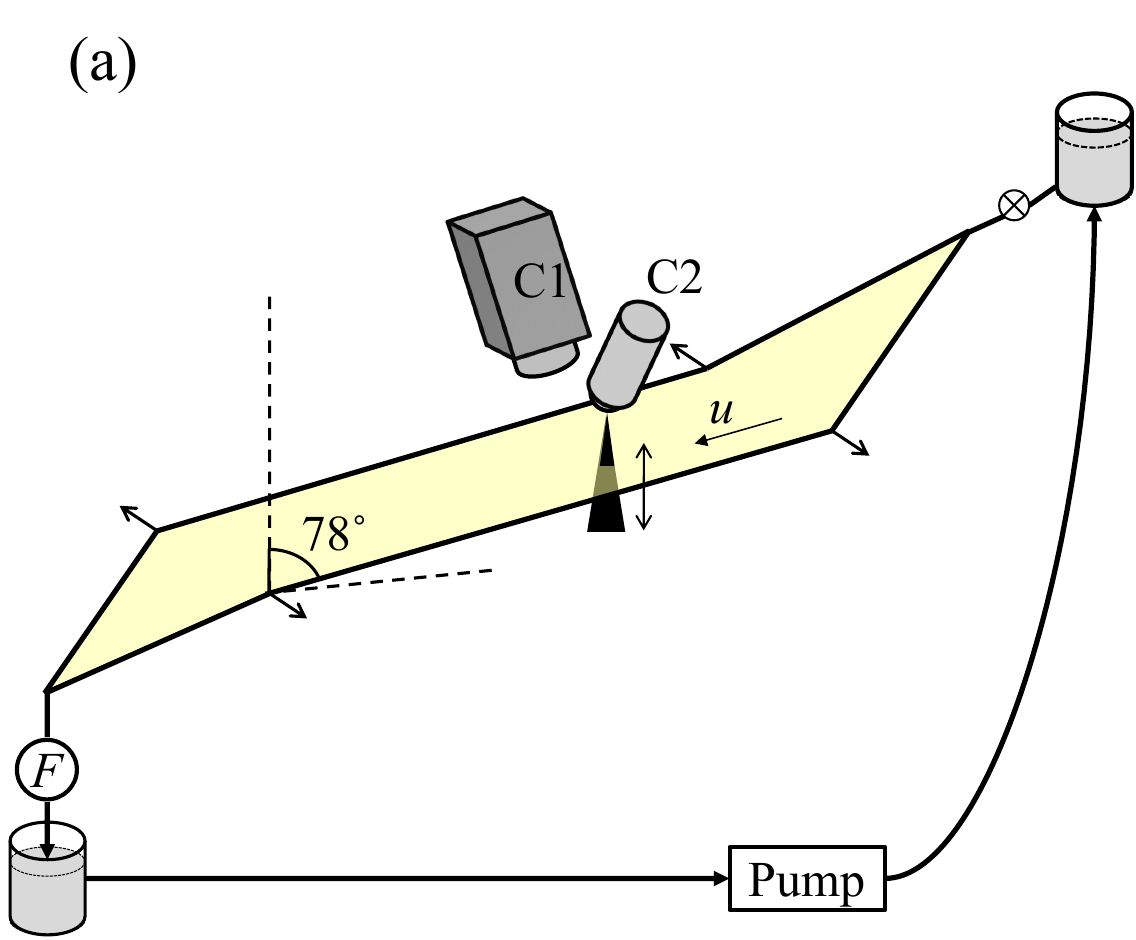}%
\includegraphics[width=0.435\textwidth]{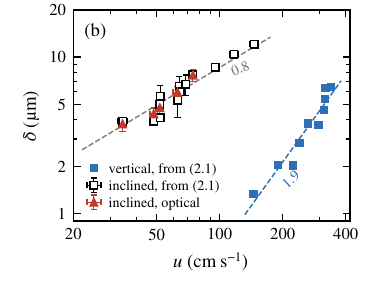}
\par
\end{centering}
\caption{
(a) The inclined channel has an angle of 78$^\circ$ from the gravity, allowing much slower flow speed $u$ than the vertical channel.
The flow rate $F$ is adjusted by manipulating a valve next to the top reservoir, and this change is distributed to the film thickness $\delta$ and $u$.
The vortex street is generated by a circular object inserted to the channel, and the flow pattern and the insertion depth are monitored by a high speed camera (C1) and a long-distance microscope (C2), respectively.
We measure $u$ by particle tracking velocimetry, and $\delta$ is determined by $\eqref{eq:thickness_flux}$. 
(b) The relationship between such measured $\delta$ and $u$ for the inclined and vertical channels. 
Optical measurements confirms that the measurement using \eqref{eq:thickness_flux} is valid.
\label{fig:exp_setup}}
\end{figure}

\ifparanumbering\paragraph{}\fi
The majority of data is measured using an inclined soap film channel in figure \ref{fig:exp_setup}(a), similar to those previously discussed \citepar{Wu:2001vq,Georgiev:2002kg,Kim:2015jp}.
Briefly, the channel consists of two flexible nylon wires, which are connected to reservoirs at the top and bottom.
The channel is about 2 m long and inclined by 78$^\circ$ from the gravity.
The top reservoir contains the soap solution, made of 2\% commercial dishsoap (P\&G Dawn) and 98\% deionized water, which is set to flow down to the channel between the two nylon wires and eventually to the bottom reservoir.
The soap solution collected at the bottom reservoir is pumped back to the top reservoir, and therefore, in principle, the soap film channel lasts indefinitely.
%The bulk kinematic viscosity $\nu=0.0126\,{\rm cm^2/s}$ is measured by Ostwald viscometer, and the Marangoni elasticity $E=22 \,\rm mN/m$ is estimated from literature \citepar{Kim:2017dn}.
The bulk kinematic viscosity $\nu=0.0126\,{\rm cm^2/s}$ is measured by Ostwald viscometer.
The Marangoni elasticity $E$ was not directly measured for this channel but is estimated to be $E\simeq22 \,\rm mN/m$ from the measurement using the same soap solution recipe \citepar{Kim:2017dn}.

%%% FIGURE 1 %%%

\ifparanumbering\paragraph{}\fi
A relatively small number of data points are measured using a vertical channel.
The vertical channel is not intended to match the statistics of the inclined channel; it occupies a different region of the $(u, \delta)$ plane, and the argument below rests on the relative placement of the two data sets rather than on the precision of individual points.
It is similar to the inclined channel described above, except in two aspects.
First, it is vertical, literally, and therefore its mean flow speed is faster.
Second, the soap solution is not recycled. 
Without the recirculation, the channel runs only one to two hours depending on the flow rate, but it does not suffer the constant dehydration due to evaporation, giving us an advantage in maintaining a constant kinematic viscosity and cleanliness.
The Marangoni elasticity has been measured for the vertical channel, confirming that $E=22\,\rm mN/m$.

\ifparanumbering\paragraph{}\fi
The mean thickness $\delta$ and flow speed $u$ of the soap film are both varied by a single control nob, the flow rate $F$.
For both channels, $F$ is varied by a valve located between the top reservoir and the channel and is measured by collecting the solution. % (inclined channel) or using a flowmeter (vertical channel).
The flow speed $u$ is measured by particle tracking velocimetry.
Then, using the continuity equation, we get
\begin{equation}
	\delta=F/(Wu),
	\label{eq:thickness_flux}
\end{equation}
where $W$ is the width of channel, which is fixed at 3.6 cm for the inclined channel and 6.9 cm for the vertical channel.

\ifparanumbering\paragraph{}\fi
The two channels distribute an increase in $F$ differently between $u$ and $\delta$.
For the inclined channel, when we vary $F$ from $0.045\,{\rm ml/s}$ to $0.64\,{\rm{ml}/s}$, $u$ varies from $0.3\,\rm m/s$ to $1.5\,\rm m/s$.
By using \eqref{eq:thickness_flux}, $\delta$ ranges from 3.5 $\mu\rm{m}$ to 13 $\mu\rm{m}$.
For the vertical channel, when we vary $F$ from $0.11\,{\rm ml/s}$ to $0.69\,{\rm{ml}/s}$, $u$ varies from $1.95\,\rm m/s$ to $3.5\,\rm m/s$.
Similarly, $\delta$ ranges from 0.8 $\mu\rm{m}$ to 2.8 $\mu\rm{m}$.
Empirically, $u\propto F^\gamma$ and $\delta\propto F^{1-\gamma}$, where $\gamma=5/9$ for the inclined channel and $\gamma=1/3$ for the vertical channel. 
We note that $\gamma=1/3$ is slightly different from $\gamma=1/4$ using a similar setup by \citebyname{Sane:2018uo}.
Then, we get $\delta \propto u^{(1-\gamma)/\gamma}$, and we find that $\delta=u^{0.8}$ for our inclined channel and that $\delta=u^{1.9}$ for our vertical channel, as seen in figure \ref{fig:exp_setup}(b).
For confirmation of the measurement using Eq. \eqref{eq:thickness_flux}, we also perform the direct measurement of $\delta$ using the transmittance of the polarized laser \citepar{Kim2010} for several data points.

\ifparanumbering\paragraph{}\fi
The vortex streets are generated by inserting a circular cylinder normal to the soap film.
For the inclined channel, we use tapered cylinders made of titanium.
The diameter $D$ of the cylinder is varied from 64 $\mu\rm{m}$ to 830 $\mu\rm{m}$ by changing the insertion depth and is measured by a long-distance microscope [C2 in figure \ref{fig:exp_setup}(a)].
The Reynolds number is defined as $\Reynolds\equiv uD/\nu$ and ranges from 50 to 550.
For the vertical channel, we use stainless steel cylinders of known diameters, ranging from 0.0184 cm to 0.107 cm.

\ifparanumbering\paragraph{}\fi
The flow patterns are visualized by fast video cameras [C1 in figure \ref{fig:exp_setup}(a)].
We use Phantom V5 from Vision Research at 1900 fps with the inclined channel, and Phantom VEO-E340L at 2700 fps with the vertical channel.
Both channels are illuminated by the low-pressure sodium lamp (589 nm), and the interferogram is recorded by high-speed video.
The thickness field is correlated with the vorticity field \citepar{Rivera:1998tw}, and therefore the interferogram captures the flow structures without any post-processing.

%%%%%%%%%%%%%%%
\ifsectioning\section{Result and Discussions}\fi
%%%%%%%%%%%%%%%

%%
\ifsectioning\subsection{The Kármán ratio}\label{sec31}\fi

\begin{figure}
\begin{centering}
\includegraphics[width=8.5cm]{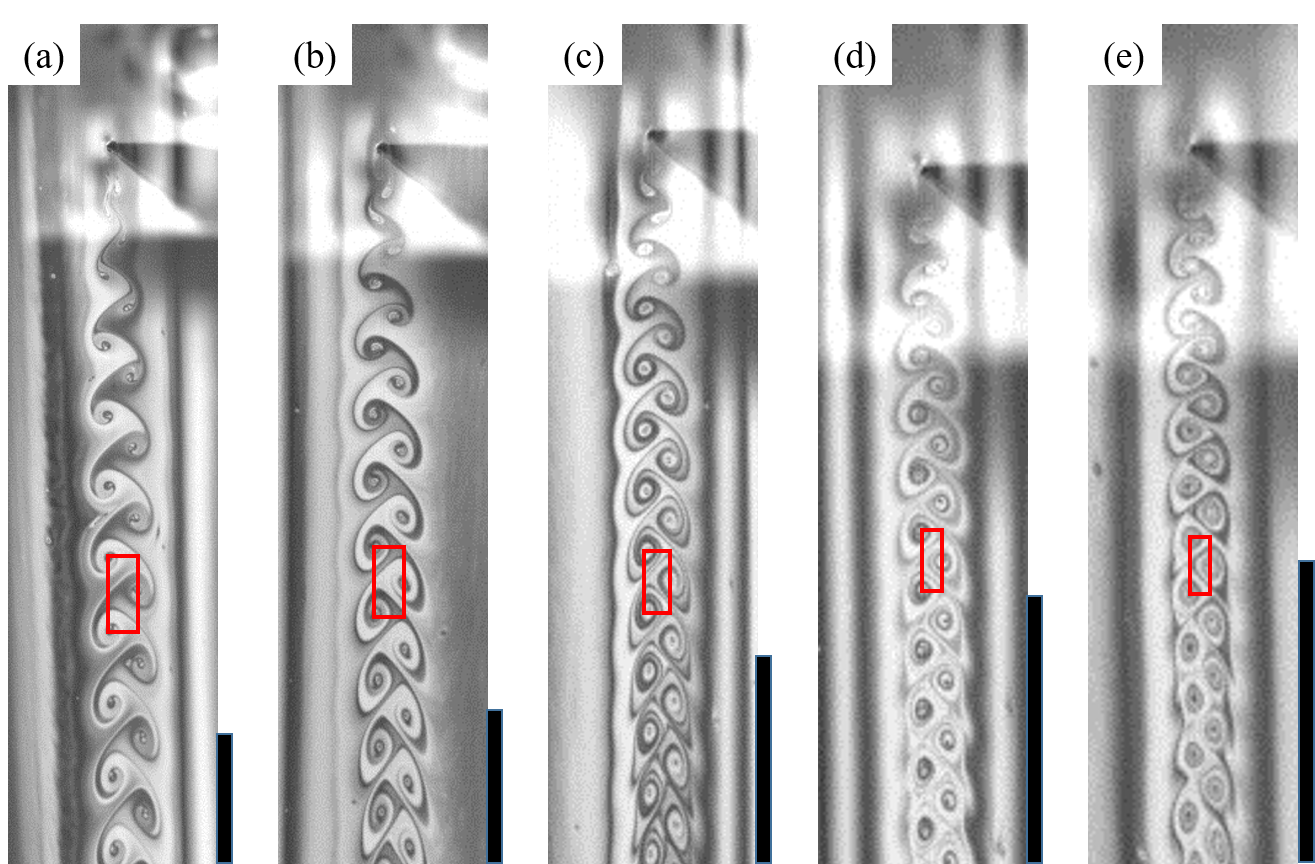}
\par
\end{centering}
\caption{
Vortex streets in the soap film at $\Reynolds \simeq 138$.
(a) $u=32\,\rm{cm/s}$ and $D=0.056\,\rm{cm}$,
(b) $u=38\,\rm{cm/s}$ and $D=0.047\,\rm{cm}$,
(c) $u=52\,\rm{cm/s}$ and $D=0.035\,\rm{cm}$,
(d) $u=64\,\rm{cm/s}$ and $D=0.027\,\rm{cm}$, and
(e) $u=77\,\rm{cm/s}$ and $D=0.024\,\rm{cm}$.
The rectangles signify the ratio $q\equiv{h}/\ell$ between the transversal and longitudinal spacing between vortices, and it is noticeably slenderer in (e) than in (a).
For dimensionless comparison, the size of images is adjusted by each respective $D$. 
Every scale bar at the lower right corner is 1 cm.
\label{fig:display_vortex_street}}
\end{figure}

\ifparanumbering\paragraph{}\fi
In figure \ref{fig:display_vortex_street}, we present a few snapshots of the vortex streets on the inclined channel.
These vortex streets are produced using different flow speed $u$ and diameter $D$ of the cylinder.
From (a) to (e), $u=32$, 38, 52, 64 and 77 [cm/s], and $D=0.056$, 0.047, 0.035, 0.027 and 0.024 [cm], respectively.
As a result, the Reynolds number $\Reynolds$ is maintained almost invariant at 138($\pm4$).
We note that images are rescaled by $D$ for non-dimensionalization; the image in figure \ref{fig:display_vortex_street}(e) ($D=0.024$ cm) is enlarged by the factor of 2.33($\approx 0.056/0.024$), compared to the image in figure \ref{fig:display_vortex_street}(a) ($D=0.056$ cm).

%%% FIGURE 2 %%%

\ifparanumbering\paragraph{}\fi
Visual inspection of the images reveals that the geometric structure of the vortex streets is not uniform, even though topologically they look alike.
This observation indicates that the principle of similitude does not apply for the soap film flows.
To quantify the difference in geometric structure, we put red rectangles to figure \ref{fig:display_vortex_street} to highlight the aspect ratio $q\equiv h/\ell$, where $h$ and $\ell$ are the transverse and the longitudinal spacing between vortices.
In general, $h$ and $\ell$ are the functions of the downstream distance $x$.
Near the cylinder, when $x<10D$, they increase as $x$ increases, but they eventually approach asymptotic values and become independent of $x$ when $x>10D$ \citepar{Roushan:2005un}.
In this work, $h$ and $\ell$ denote the asymptotic value and independent of $x$. 
This ratio $q$, known as Kármán's ratio, has been discussed by Th. von Kármán in 1911 using the point vortex model \citepar{vonKarman:1911vi, vonKarman_reprint, Saffman}.
The stability analysis implies that only vortex streets with $q=(1/\pi) \tanh^{-1}({1}/{\sqrt{2}})\approx0.28$ are stable, assuming zero viscosity and infinitesimal vortices.

\begin{figure}
\begin{centering}
\includegraphics[width=\textwidth]{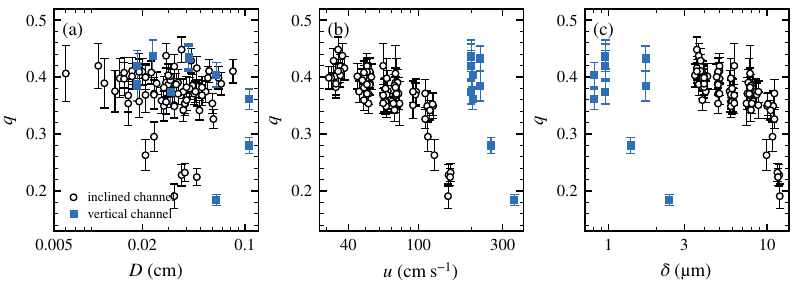}
\par
\end{centering}
\caption{
The ratio $q$ between the transversal spacing $h$ and the longitudinal spacing $\ell$ between vortices, plotted with respect to (a) the object size $D$, (b) the mean flow speed $u$, and (c) the film thickness $\delta$.
Data from the inclined channel (open black circles) and from the vertical channel (closed blue squares) are shown.
No direct correlation between $q$ and $D$ is observed. 
However, $q$ is strongly correlated with $u$ and $\delta$ if they are grouped by the channel, but the results from different channels are not compatible.
%The horizontal dash lines show the prediction of the point vortex model $q=0.28$.
\label{fig:q_scaling}}
\end{figure}

\ifparanumbering\paragraph{}\fi
According to our measurements, $q$ depends on none of $D$, $u$, $\delta$ alone, as shown in figures \ref{fig:q_scaling}(a-c).
In the figures, $q$ measured from both inclined and vertical channels under various experimental conditions, $D$ and $F\propto u^\gamma \delta^{1-\gamma}$, are displayed. 
In figure \ref{fig:q_scaling}(a), it is shown that $D$ is not a determining factor of $q$, regardless of the channels' orientation.
While $D$ is varied over a decade, no statistically significant correlation is observed between $q$ and $D$. 
However, partial trends are observed when $q$ is plotted with respect to $u$ or $\delta$. 
In figures \ref{fig:q_scaling}(b) and (c), $q$ is observed to be decreasing as $u$ and $\delta$ increases when data are grouped by channels. 
These observations collectively indicate that none of $D$, $u$ or $\delta$ solely determine $q$, and, as a corollary, $\Reynolds$ alone does not determine $q$ either.

%%% FIGURE 3 %%%

\ifsectioning\subsection{Effect of viscosity}\label{sec32}\fi

\ifparanumbering\paragraph{}\fi
Figure \ref{fig:q_scaling}(c) indicates that the effect of film viscosity on $q$ is not substantial.
The viscosity of soap film is described by the Trapeznikov relation $\eta_f=2\eta_s+\eta_b\delta$, where $\eta_b$ is the dynamic viscosity of the bulk liquid and $\eta_s$ is the contribution of the surface viscosity \citepar{Trapeznikov,Prasad:2009jj}.
By dividing the relationship by the density, we get the kinematic viscosity relation $\nu_f=2\nu_s/\delta+\nu_b$.
Therefore the film viscosity $\nu_f$ depends only on $\delta$. 
However, our data show that $\delta$ is not a sole factor for the vortex geometry.

\ifparanumbering\paragraph{}\fi
Even if we limit our discussion of film viscosity to a single channel, for example the inclined one, the variation of film viscosity is limited to only a few percent.
Using the measurements of \citebyname{Prasad:2009bk} that $\eta_s\sim 0.5 \,\rm nPa\, m \, s$, we get $\nu_s/\delta\lesssim0.0014 \,\rm cm^2/s$ at $\delta=3.5\,\rm\mu{m}$.
This number by itself is approximately 10\% of the bulk viscosity $\nu_b=0.0126\,\rm cm^2/s$. 
%Considering that the ranges of variations of $\ell$ and $h$ are much greater than 10\%, it is unlikely that our observation of variations of $h_0$ and $\ell_0$ is due to the viscosity variation. 
Moreover, we recall the mathematical investigation by \citebyname{Hooker:1936tz}, where he concluded that ``pure spread of eddies causes no widening of the street".

\ifparanumbering\paragraph{}\fi
To summarize, the flow speed dependence in soap film experiments has often been attributed to the accompanying change in viscosity \citepar{Vorobieff:1999wn}, however, our data show that viscosity, which depends on $\delta$ alone, cannot account for the observed variation of $q$.

\ifsectioning\subsection{Measurements of $h$ and $\ell$}\label{sec33}\fi

\begin{figure}
\begin{centering}
\includegraphics[width=\textwidth]{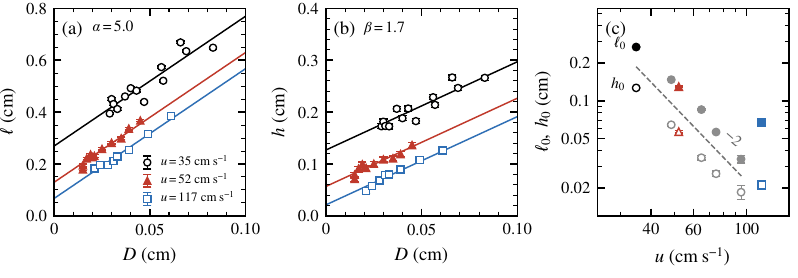}
\par
\end{centering}
\caption{
The measurement of (a) $\ell$ and (b) $h$ plotted with respect to $D$.
Data are shown to follow the linear relationship between $h$ and $D$ and $\ell$ and $D$, but the linearity is observed only when data are grouped by $u$.
Here, three cases of $u$, 35$\pm$3 cm/s (open black circles), 52$\pm$3 cm/s (closed red triangles), and 117$\pm$6 cm/s (open blue squares) are shown.
The proportionality constant $\alpha=5.0$ and $\beta=1.7$ are irrespective of $u$, but the intercept $h_0$ and $\ell_0$ depends on $u$.
\label{fig:l_and_h}}
\end{figure}

\ifparanumbering\paragraph{}\fi
While the Kármán ratio $q$ is independent of $D$, $h$ and $\ell$ individually depend on $D$ in a linear manner.
In figures \ref{fig:l_and_h}(a,b), the measurement of $h$ and $\ell$ are presented with respect to $D$.
The measurement shows that both $h$ and $\ell$ are linearly proportional to $D$, which is not particularly surprising because the linearity has been already reported in literature \citepar{Roushan:2005un,Kim:2015jp}.
However, this linearity is observed only when the data are grouped by the flow speed $u$.
From figures \ref{fig:l_and_h}(a,b), it is implied that the scaling relation follows the form 
\ifjfm\begin{subeqnarray}\fi
\ifaps\begin{eqnarray}\fi
h&=&h_0(u)+\beta D,
\\
\ell&=&\ell_0(u)+\alpha D, 
\label{eq:linearity}
\ifjfm\end{subeqnarray}\fi
\ifaps\end{eqnarray}\fi
where $\alpha=5.0$ and $\beta=1.7$ are proportionality constants independent of $u$, and $\ell_0$ and $h_0$ are intercepts.% that are decreasing functions of $u$ as seen in figure \ref{fig:l_and_h}(c).

%%% FIGURE 4 %%%

\ifparanumbering\paragraph{}\fi
The two intercepts, $h_0$ and $\ell_0$, have strong dependence on $u$ as shown in figure \ref{fig:l_and_h}(c). 
Nevertheless the variation of $u$ is less than a decade, there is a clear trend at relatively slower flow speed.
For $u<100\,\rm cm/s$, both $h_0$ and $\ell_0$ are inversely proportional to $u^2$, i.e. $h_0\propto u^{-2}$ and $\ell_0\propto u^{-2}$.
It seems that the highest-speed datum deviates from the trend, but unfortunately, we are unable to produce additional data points for $u>117\,\rm cm/s$ for technical reason.

\ifparanumbering\paragraph{}\fi
Our data indicate that the variation of intercepts may not be attributed to the effect of meniscus. 
Every object inserted through a soap film may need its size to be corrected with the formation of meniscus around it. % \citepar{Kim.2023tim}.
The size of meniscus $d$ depends both on $u$ and $D$. 
For its dependence on $D$, we recall the study by \citebyname{Kim.2023tim} where experimental observation indicates $d \propto D$ for small $D<0.05\,\rm cm$, overlapping with the experimental range of current study.
Therefore, the $D$-dependence of $d$ only appear in $\alpha$ and $\beta$ in \eqref{eq:linearity} but has no effect on $h_0$ and $\ell_0$.
In contrast, the $u$-dependence of $d$ is linked to $h_0$ and $\ell_0$. 
The study on menisci formation by \citebyname{clanet2002} shows that the meniscus height is proportional to the square root of its characteristic time of establishment.
When this time scale is assumed to be $D/u$, it follows that $d\propto u^{-\frac{1}{2}}$, but this exponent is much more gradual than -2, as observed in our experiments.

\ifparanumbering\paragraph{}\fi
Before we move on to the next discussion, we estimate a few $q$ values using the parameters of \eqref{eq:linearity}.
When $D\rightarrow 0$, the ratio $q$ approaches $q_0\equiv h_0/\ell_0$, which is independent of $u$. 
Also when $u\rightarrow 0$, $h\rightarrow h_0$ and $\ell\rightarrow\ell_0$, therefore it is expected that $q\rightarrow h_0/\ell_0$.
On the other hand, when $D\rightarrow\infty$, $q$ approaches $\beta/\alpha=0.34$.
Also when $u\rightarrow \infty$, $q\rightarrow \beta/\alpha$ because the intercepts are less important. 
Therefore, the linear model allows us to estimate approximate range of $q$, i.e. $0.34<q<0.48$. 
This may seem to be contradictory with our observation at the first glance because we actually observe data points outside this range.
However, $h_0$ and $\ell_0$ being inversely proportional to $u^{-2}$ is an observation within a short range of $u$.
Therefore, this calculation should be accepted as statistically averaged range, and the linear model does not fully capture our experimental observation.

\ifparanumbering\paragraph{}\fi
To summarize, our results clearly show that $\Reynolds$ is not a single parameter of soap film channel.
When the linear models in \eqref{eq:linearity} are non-dimensionalized, we get $h/D = h_0(u)/D+\beta$ and $ \ell/D= \ell_0(u)/D + \alpha$.
If $h_0$ and $\ell_0$ were proportional to $u^{-1}$, the first terms in the right-hand sides are proportional to $\Reynolds^{-1}$ alone.
However, it has an additional $u^{-1}$ dependence and cannot be rescaled uniformly under the principle of similitude.

\ifsectioning\subsection{Collapse under $u\sqrt{\delta}$}\label{sec34}\fi

\begin{figure}
\begin{centering}
\includegraphics[width=\textwidth]{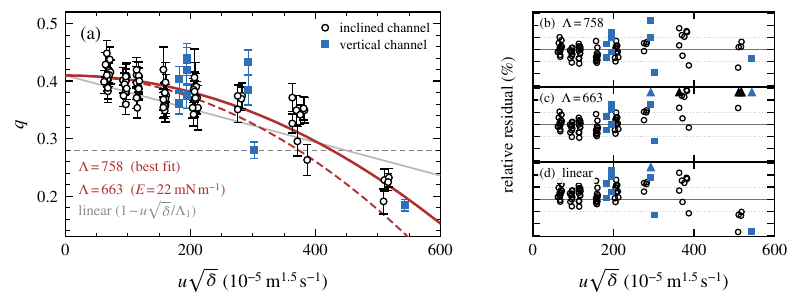}
\par
\end{centering}
\caption{
(a) The measurement of $q$ from two channels collapse into a single curve when they are plotted with respect to $u\sqrt\delta$.
While the linear model, $q/q_0=(1-u\sqrt\delta/\Lambda_1)$, fails capture the trend, the quadratic model, $q/q_0 = 1- u^2\delta /\Lambda^2$ fits data well.
The best fit parameter is $\Lambda=758\times 10^{-5} \,\rm m^{1.5}s^{-1}$, but $\Lambda=663\times 10^{-5} \,\rm m^{1.5}s^{-1}$, at which $u^2\delta=M^2$, is also displayed.
Horizontal line marks $q=0.28$ from the point vortex model.
(b,c,d) The relative residuals are displayed for three curves in (a). 
The range of each ordinate is $\pm$30\%, with the horizontal dashes show $\pm10\%$.
Triangular symbols indicate the presence of data points out of the displayed range.
\label{fig:collapse}}
\end{figure}

\ifparanumbering\paragraph{}\fi
Although neither $u$ nor $\delta$ alone determines $q$, the data from both channels collapse onto a single curve when plotted with respect to the combination $u\sqrt\delta$, as seen in figure \ref{fig:collapse}.
Although $u$ and $\delta$ of two channels differ by factor $\sim 2$ and $\sim 6$ respectively, the collapse of data is considerable.
The use of $u\sqrt\delta$ is directly motivated from the observation that data from two channels are separated in opposite direction in figures \ref{fig:q_scaling}(b) and (c).
However, this combination is not arbitrary as implied by the dimensional analysis.
When $D$ and $\nu$ are considered to be irrelevant variables as we discussed earlier, the remaining variables are $u$, $\delta$, $E$, and $\rho$.
Then Buckingham's Pi theorem allows one dimensionless number $\Pi=(\rho/E) u^2 \delta $. 

%%% FIGURE 5 %%%

\ifparanumbering\paragraph{}\fi
We note that $u\sqrt\delta$ is another form of Mach-like number in flowing soap films. 
In soap film flows, the Marangoni elastic wave defines the speed at which the thickness fluctuation propagates and is expressed as $\vm=\sqrt{2E/\rho \delta}$.
According to the gasdynamic analog of soap film flows, the thickness of soap film flows behaves like the density of compressible gas flow, and therefore the Marangoni elastic wave speed is analogous to the acoustic speed.
Then, it is natural to define a Mach-like number $M\equiv u/\vm$.
Using the definition of $\vm$, we get $M=u\sqrt{\delta}/\Lambda$, where $\Lambda=\sqrt{2E/\rho}$, and $\Pi=2M^2$.

\ifparanumbering\paragraph{}\fi
The Mach-like number $M$ is closely related to the compressibility of soap film flows.
The momentum equation of steady soap film flows is expressed as \citepar{Kim:2017dn}
\begin{equation}
\rho u du = 2d\sigma/\delta.
\label{eq:momentum}
\end{equation}
Using the definition of Marangoni elasticity $E=-\delta({d\sigma}/{d\delta})$, it follows that 
\begin{equation}
({d\delta}/{\delta})=-M^2 ({du}/{u}).
\label{eq:compressibility}
\end{equation}
Therefore $M^2$ is considered as a measure of compressibility; as $M$ increases, the compressibility increases in a quadratic manner.

\ifparanumbering\paragraph{}\fi
We find that $q$ depends on the compressibility of the channel; as seen in figure \ref{fig:collapse}(a), the data follows the form
\begin{equation}
\frac{q}{q_0}=  1- \left(\frac{u\sqrt\delta}{\Lambda}\right)^2.
\label{eq:fitting}
\end{equation}
From the regression analysis, $\Lambda=758 \times 10^{-5} \,\rm m^{1.5}s^{-1}$ is acquired, which may imply $E=28\,\rm mN/m$.
Using our established value $E=22 \,\rm mN/m$, $\Lambda=\sqrt{2E/\rho}=663\times 10^{-5}\,\rm m^{1.5} s^{-1}$ is expected, however the best fit value is approximately 14\% higher.
Using the currently available data, it is difficult to call whether $q/q_0\sim 1-M^2$, $q/q_0\sim 1-kM^2$ ($k$ is a coefficient), or $q/q_0\sim 1/(1+M^2)$ best describe data because measurement is extremely difficult at large $M$.
We could obtain only a few data points at large $M$, but they have high leverage in the regression analysis.
However, the linear relation, both $1-M$ and $1/(1+M)$ scalings, certainly goes out of sync with data, as seen in the relative residuals of some models provided in figures \ref{fig:collapse}(b-d).

\ifparanumbering\paragraph\fi
The observed $M^2$-dependence is also consistent with the usual weak compressibility expansion. 
In the Janzen-Rayleigh expansion \citepar{Liepmann}, compressible corrections to an incompressible base flow appear at $\mathcal{O}(M^2)$. 
For example, \citebyname{Crowdy:2017cs} showed that the propagation speed of a weakly compressible von Kármán vortex street acquires an $\mathcal{O}(M^2)$ correction. 
While their result concerns the speed rather than the street geometry, it provides an independent theoretical precedent for an $M^2$-order modification of vortex street dynamics.
On the other hand, the linear scaling generally appears in Doppler effect in a form $\frac{(1-M_o)}{(1+M_s)}$, where the subscripts $o$ and $s$ stand for `observer' and `source'.
Our observation that $q/q_0=1-\mathcal{O}(M^2)$ indicates that the primary physics behind the breakdown of $\Reynolds$ scaling in soap film is due to the compressibility effect rather than the compression of wave-like motion.

%%%%%%%%%%%%%%%
\ifsectioning\section{Conclusion}\label{sec4}\fi
%%%%%%%%%%%%%%%

\ifparanumbering\paragraph{}\fi
To summarize, we have investigated the effect of mean flow speed on the flow structure in soap film flows.
We quantified the geometric structure of vortex street behind a circular object with the Kármán ratio, the ratio between the transverse to the longitudinal spacing between vortices, which we observe decreasing as the flow speed increases, even though the Reynolds number was kept invariant. 
We inspected vortex streets generated by using various object sizes, mean flow speed $u$, film thickness $\delta$, and the channel inclination angle, and we find that the Kármán ratio $q$ variation cannot be accounted for by a single independent variable.
Instead, our data collapse onto a single curve when plotted with respect to $u\sqrt\delta$.

\ifparanumbering\paragraph{}\fi
Physically, the combination $u\sqrt\delta$ can be interpreted as a Mach-like number $M\equiv u/\vm$, where $\vm$ is the Marangoni elastic wave speed.
Analytically, $q/q_0= 1-(u\sqrt\delta/\Lambda)^2$, where $\Lambda$ is a fitting parameter, matches data well, even though $\Lambda$ from the best fit is 14\% larger than theoretically expected.

\ifparanumbering\paragraph{}\fi
Our study indicates that it is the compressibility that alters flow structure. 
This is supported by the emergence of $\mathcal{O}(M^2)$ as a leading term, which is a hallmark of compressible flow.
For many years, the flow speed of soap film channels has been believed to be kept constant because of viscosity, but our study shows that the compressibility is the primary reason.

\ifparanumbering\paragraph{}\fi
To conclude, the soap film channel should be considered as a model system of two-dimensional Navier-Stokes equation only under a specific context.

%%%%%%%%%%%%%%%
\ifsectioning\section*{Acknowledgement}\fi
%%%%%%%%%%%%%%%

This work was supported by the National Research Foundation of Korea (NRF) grant, funded by the Korean government (MSIT) (No. 2021R1C1C1010158).

\bibliography{soapfilmcaprice2026}

\end{document}